\documentclass[icps3]{svjour}
\usepackage{graphicx}
\begin{document}

\title{Resonant tunneling through zero-dimensional impurity states:
Effects of a finite temperature}

\titlerunning{Effects of finite temperature in tunneling}

\author{P.~K\"onig,
      U.~Zeitler, J.~K\"onemann, T.~Schmidt,
       R.~J.~Haug}

\authorrunning{P.~K\"onig et al.}

\institute{Institut f\"ur Festk\"orperphysik, 
  Universit\"at Hannover, Appelstra{\ss}e 2, 30167 Hannover, Germany\\
  \email{zeitler@nano.uni-hannover.de}}

\maketitle

\begin{abstract}
  We have performed temperature dependent tunneling experiments 
  through a single impurity in an asymmetric vertical double 
  barrier tunneling structure.
  In particular in the charging direction we observe at zero magnetic field 
  a clear shift in the onset voltage of the resonant tunneling current through
  the impurity. With a magnetic field applied
  the shift starts to disappear. The experimental observations are
  explained in terms of resonant tunneling through a spin 
  degenerate impurity level.
\end{abstract}

Resonant tunneling experiments through a single impurity~\cite{Read}
mainly reflect the energetic position of an impurity level
and the Fermi-Dirac distribution in the emitter.   
In order to refine the concept of such tunneling processes,
in particular as far as the temperature dependence is concerned,
we investigated an asymmetric double-barrier resonant-tunneling device
(DBRTD) consisting of a 10 nm GaAs quantum well (QW) and two 5~nm and 8~nm wide
Al$_{0.3}$Ga$_{0.7}$As barriers. The structure is sandwiched between
highly doped GaAs electrodes 
(Si-doped with $n_{Si} = 4 \times 10^{17}$~cm$^{-3}$)
separated from the barriers by a 7 nm thick nominally undoped 
GaAs spacer. From this wafer we processed a vertical tunneling
diode with a mesa diameter of 2~$\mu$m. When applying a bias
voltage $V$ between the top and the bottom electrode 
the sample displays the normal behavior of a resonant
tunneling diode with pronounced current peaks at $V_r^+ = 82~$mV and
$V_r^{-} = -183$~mV. For lower voltages additional small current steps 
are observed. We assign them to tunneling through zero-dimensional
impurity levels originating from donor atoms unintentionally present in
the GaAs QW~\cite{Read,Bumbel}. 
In this work we concentrate on the first current step due to resonant tunneling
through the energetically lowest lying impurity state.
We will show that the spin degeneracy of this ground state
and the Coulomb blockade causes a shift of the step position
as a function of temperature~\cite{note}.

\begin{figure}[t]
  \centerline{\includegraphics[width=0.4\hsize]{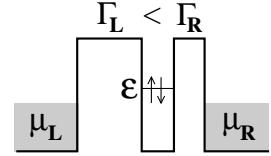}}
  \caption{Energy-level scheme showing the two-fold spin-degenerate 
  ground state of an impurity at energy $\epsilon$ coupled to 
  electrodes with chemical potentials  $\mu_L$ and $\mu_R$. 
   The tunneling rates through the two barriers are  
  $\Gamma_L$ and $\Gamma_R$. }
  \label{Sketch}
\end{figure}

For a theoretical description of the effects observed
we regard the situation as sketched in Fig.~\ref{Sketch}.
A zero-dimensional impurity with a spin-degenerate ground state
$\epsilon$ is situated inside the central well of the DBRTD.
It is coupled via two tunneling barriers (characterized by tunneling rates 
$\Gamma_L < \Gamma_R$) to 
three-dimensional reservoirs with chemical potentials $\mu_L$ and $\mu_R$.
Applying a finite transport voltage $V$ to the structure induces a difference
in the chemical potentials, $eV = \mu_L- \mu_R$.
A resonant current through the impurity shows up as a current step
at a voltage $V_+ > 0$ 
(or $V_- <0$, respectively) when $\mu_L$ (or $\mu_R$, respectively) 
equals $\epsilon$~\cite{note}.
Two possible spin states can be occupied by an electron during
a tunneling event. However, Coulomb blockade prohibits  
simultaneous occupancy of both states at the same time. 
With this condition the tunneling current $I$ through the impurity 
can be calculated as~\cite{Schoeller}:
\begin{equation} 
I  = 2 e \; \Gamma_L \Gamma_R \;
       \frac{f_L (\epsilon) - f_R (\epsilon)}
       { \Gamma_L + \Gamma_R  + \Gamma_L f_L(\epsilon)
           + \Gamma_R f_R(\epsilon) }
\end{equation}
where $f_L (\epsilon)$ and $f_R (\epsilon)$ are the 
Fermi-Dirac distributions in the left and right reservoirs.

For a finite bias $|eV| \gg kT$ 
we have $f_R \ll 1$ for $V >0$ (and $f_L \ll 1$ for $V<0$) and 
the current as a function of an applied bias $V$ simplifies to:
\begin{eqnarray} 
I  =  2 e \Gamma_L \Gamma_R 
       \frac{f_L (\alpha_L e (V_+-V) )}
       { \Gamma_L + \Gamma_R + \Gamma_L f_L(\alpha_L e (V_+-V ))}
       ~~(V > 0) \nonumber \\[-0.5em]
       \hspace*{5cm}(2a) \nonumber\\[0.5em]
I =   - 2 e   \Gamma_L \Gamma_R 
       \frac{f_R (\alpha_R e (V-V_-) )} 
       {\Gamma_L + \Gamma_R  + \Gamma_R f_R(\alpha_R e (V-V_-))}
       ~~(V <0)  \nonumber \\[-0.5em]   
        \hspace*{5cm}(2b) \nonumber
\end{eqnarray}
The prefactors $\alpha_L$ and $\alpha_R$ account for the fact that
only a part of the bias voltage drops between the emitter reservoir
and the impurity, see also below.  

\begin{figure}[t]  
  \centerline{\includegraphics[width=0.85\hsize,angle=0]{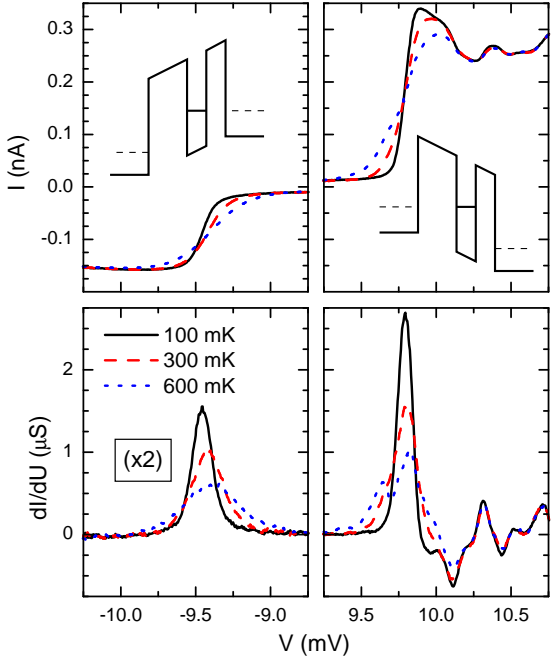}}
   \caption{Current voltage characteristics (top) and differential conductance
           (bottom) of a vertical asymmetric DBRTD
           for different temperatures.
           The steps are due to resonant tunneling through a single  impurity.
           The left panels represent the charging direction, the right panels
           show the non-charging direction at positive bias voltages.
           The insets sketch the energy-level schemes of the structure at
           the step voltages $V_{-}$ and $V_+$ for the two bias directions.}  
  \label{IU}
\end{figure}

An experimental example for the $I$-$V$ characteristic of our
sample is shown in Fig.~\ref{IU}. Clear current steps due to resonant 
tunneling through the energetically lowest lying impurity level are observed
for both bias directions in the two top figures. For positive
bias (right panel) the electrons tunnel into the impurity through 
the thicker barrier $\Gamma_L$
and leave it through the thinner one $\Gamma_R$.
The impurity is mostly empty and both degenerate states
of $\epsilon$ are available for tunneling (non-charging 
direction). For negative bias 
the tunneling current is limited by the small tunneling $\Gamma_L$
for electrons leaving the impurity. As a consequence, one of the two states
of $\epsilon$ is mostly occupied and Coulomb blockade suppresses
a simultaneous tunneling event through the other 
spin state (charging direction). Therefore,
the resonant current is smaller than in 
the non-charging direction. This current suppression 
follows from Eqs.~(2a) and (2b) which predict for
$\Gamma_R \gg \Gamma_L$ and $T=0$ current steps  
$\Delta I^{+} = 2e \Gamma_L$ in the non-charging direction and   
$\Delta I^{-} = -e \Gamma_L$ in the charging direction.

\begin{figure}[t]
  \centerline{\includegraphics[width=0.7\hsize,angle=0]{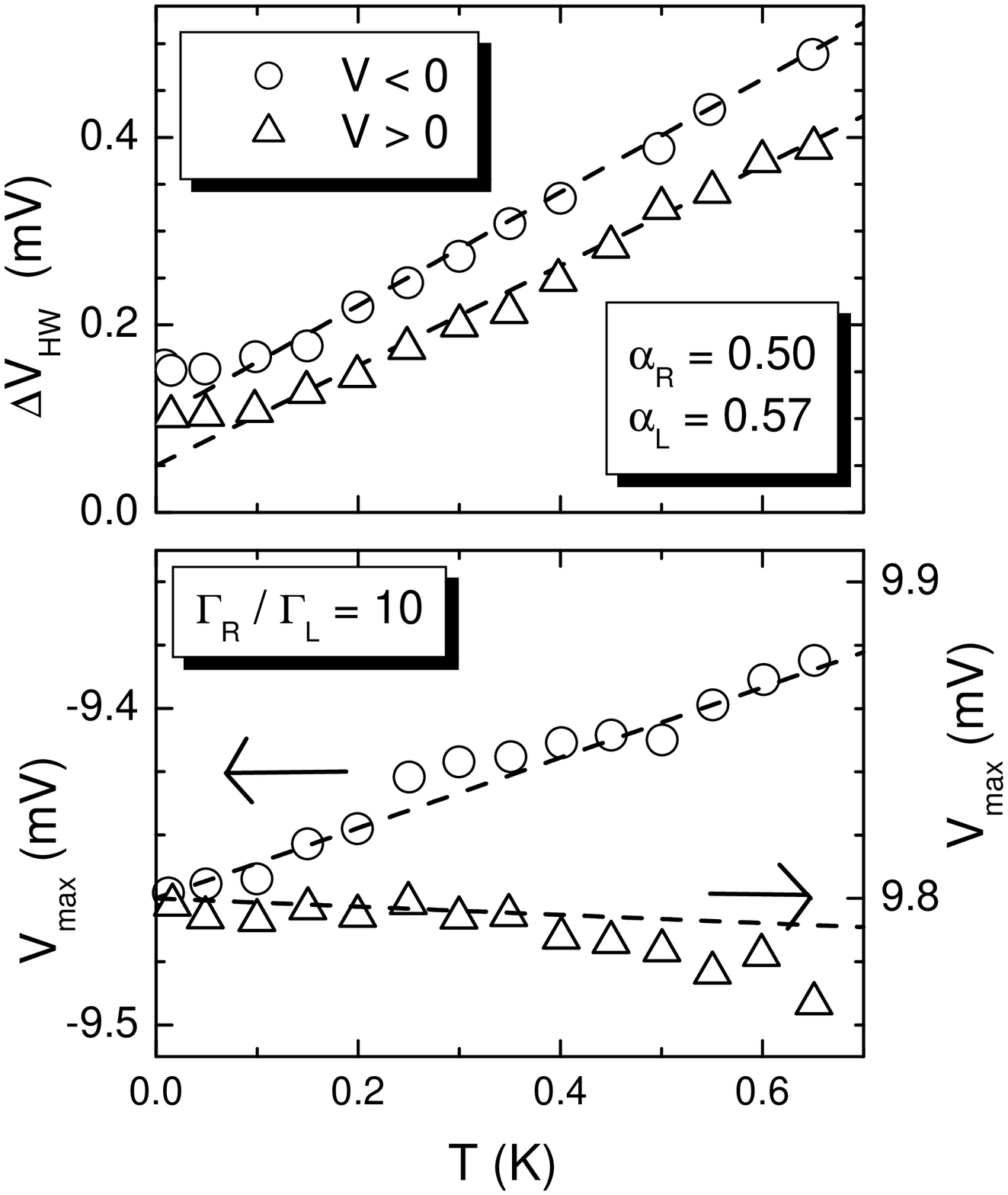}}
  \caption{Top: Temperature dependence of the half-width $\Delta V_{HW}$
            of the conductance peak for the two bias directions.
            The lines show fits for lever factors.
           \protect{\newline}
           Bottom: Experimental temperature dependent shift of the 
           conductance-peak position for both bias directions (symbols)
           compared to the theoretical behavior using 
           $\Gamma_R / \Gamma_L = 10$ (dashed lines).
          }
  \label{alf-pos}
\end{figure}

Effects of temperature can be visualized more clearly when plotting the
differential conductance $G=dI/dV$ as a function of bias voltage.
The current steps show up as peaks in $G$. Fitting its half width
$\Delta V_{HW}$ as a function of temperature to Eqs.~(2a) and (2b) 
allows to determine the lever factors $\alpha_L$ and $\alpha_R$,
see Fig.~\ref{alf-pos}, top panels. For both bias directions
we find $\alpha \approx 0.5$ indicating that the impurity is approximately
situated half way between the two reservoirs.

Eqs.~(2a) and (2b) also predict 
that the maximal conductance at finite 
temperature is not always observed at the voltage where the chemical
potential in the source (i.e.~$\mu_L$ for
positive bias and $\mu_R$ for negative bias) equals $\epsilon$. In particular
in the charging direction the step voltage shifts to a lower absolute
value when $T$ is increased. This is indeed observed experimentally and
shown in the left bottom curve of Fig.~\ref{alf-pos}. The solid line shows
a theoretical fit as expected from Eq.~(2b) and yields 
$\Gamma_R/\Gamma_L = 10$ in reasonable agreement with the expected 
tunneling rates through the two barriers. For the non-charging
direction the shift observed is negligible and consistent 
with  $\Gamma_R/\Gamma_L = 10$.

When applying a magnetic field the Zeeman effect lifts the spin-degeneracy 
of $\epsilon$. As a consequence the current step in the SET
direction splits-into two steps corresponding to tunneling of 
electrons with a different spin orientation~\cite{Desh,PeterSpin}. From
the magnetic field dependence of this split we extract
a Land\'e factor $g^* = -0.14$ for the impurity ground state~\cite{PeterSpin}.
In the charging direction Coulomb blockade still prohibits
simultaneous tunneling through both energy levels and only
one current step is observed. However, since the degeneracy
is lifted the shift of the step voltage is no more observed 
as long as $3.5 k_BT > g^* \mu_B B$. 
For $3.5 k_BT < g^*\mu_B B$ the energy splitting
of the ground state $\epsilon$ is no more resolved and $\epsilon$
can be again regarded as a virtually degenerate level. 
Finally it is worthwhile remarking that the effects described in
this paper also influence the spin splitting of the current steps
in the SET direction and a direct determination of $g^*$ is only
possible if $3.5 k_BT < g^* \mu_B B$. 

In conclusion we have shown that the actual voltage position 
of a resonant current step
through a zero-dimensional state is not only given by the energetic
position of this state but can also be strongly influenced
by temperature.


\begin{thebibliography}{9}
\bibitem{note}
Here and in the following we define the step voltages $V_+$ and $V_-$ 
as the voltages where the current steps exhibit the largest slope.
\end{thebibliography}
\end{document}